\documentclass[12pt,preprint]{aastex}

\usepackage{graphicx,rotate}
\usepackage{natbib}
\usepackage{latexsym,amssymb,verbatim}

\shorttitle{Hot core models}
\shortauthors{Bayet et al.}

\begin{document}


\title{Molecular tracers of high mass star-formation in external galaxies}


\author{E. Bayet\altaffilmark{1}, S. Viti\altaffilmark{1}, D.A. Williams\altaffilmark{1}
        and J.M.C. Rawlings\altaffilmark{1}}

\email{eb@star.ucl.ac.uk}


\altaffiltext{1}{Department of Physics and Astronomy, University
College London, Gower Street,
       London WC1E 6BT, UK.}


\begin{abstract}
Hot core molecules should be detectable in external active galaxies
out to high redshift. We present here a detailed study of the
chemistry of star-forming regions under physical conditions that
differ significantly from those likely to be appropriate in the Milky
Way Galaxy. We examine, in particular, the trends in molecular
abundances as a function of time with respect to changes in the
relevant physical parameters. These parameters include metallicity,
dust:gas mass ratio, the H$_{2}$ formation rate, relative initial
elemental abundances, the cosmic ray ionization rate, and the
temperature of hot cores. These trends indicate how different tracers
provide information on the physical conditions and on evolutionary
age. We identify hot core tracers for several observed galaxies that
are considered to represent spirals, active galaxies, low-metallicity
galaxies, and high-redshift galaxies. Even in low-metallicity
examples, many potential molecular tracers should be present at
levels high enough to allow unresolved detection of active galaxies
at high redshift containing large numbers of hot cores.
\end{abstract}





\section{Introduction}\label{sec:intro}

Hot cores are commonly observed in regions of massive star formation
in the Galaxy, through their molecular line emissions. These hot
cores correspond to a relatively small, dense, warm zone of gas
surrounding a newly-formed star. They show a characteristic
chemistry, distinct from that of molecular clouds in more quiescent
regions of the interstellar medium \citep{Walm93}. Hot cores are rich
in relatively large organic molecules, some are products of surface
chemistry. This characteristic chemistry is generally attributed to a
prolonged evolution of the gas and dust at very low temperatures,
high densities, and large visual extinctions, and it is therefore
inferred that the hot cores represent a sample of material from the
pre-stellar gas that was not incorporated into the newly-formed star
\citep{Walm93}.

While stellar spectra give information on local elemental abundances
\emph{after} star formation has occurred, molecular line intensities
from hot cores can - with the use of an appropriate hot core model -
give information on the local elemental abundances \emph{before} star
formation occurred. Such an opportunity may be particularly useful in
external galaxies where the metallicity and other physical parameters
may be considerably different from those in the Milky Way Galaxy.
Hence, \citet{Lint05} suggested that hot cores could be used as
probes of high redshift galaxies in which the rate of star formation
and the hot core population are much larger than in the Milky way
Galaxy. Ideally, one would be able to distinguish between different
models of stellar evolution for the yields of metallicity from
zero-metallicity stars. \citet{Lint05} showed that if hot cores in a
high redshift galaxy were sufficiently numerous, then molecular line
emission from a large population of unresolved hot cores in high
redshift galaxies should be detectable. However, there exist at
present no models of the chemistry of hot cores in the physical
conditions likely to pertain in high redshift galaxies.

Therefore, our intention in this work is to calculate the expected
chemistry arising in hot cores in galaxies which have a wide range of
physical conditions that may be substantially different to those
found in the Milky Way Galaxy. The physical parameters that
significantly affect the chemistry include gas density and
temperature, metallicity, dust : gas ratio, cosmic ray flux and its
ionization rate, interstellar radiation field, etc. Of course, hot
cores are only one component giving rise to molecular line emission
in external galaxies. Other sources of molecular line emission are
photon dominated regions (PDRs) and cool molecular clouds (see, e.g.
\citealt{Garc06,Garc07}). However, hot cores have a characteristic
chemistry which should give rise to a particular line emission
signature. We shall in future work examine the emission signature
from other types of sources.

We use as the basis of our computations the hot core model of
\citet{Viti99} which was developed for the description of hot cores
in the Milky Way. In the present work we are particularly concerned
with galaxies at high redshift in which frequent massive star
formation and the most intense hot core molecular line emissions
occur. The rise time of the stellar temperature from that of the cold
gas to the Main Sequence is relatively short \citep{Bern96, Hans98} and, for
present purposes, we shall regard this as instantaneous (unlike the
treatment of \citealt{Viti04}, in which the evolving chemistry of the
hot core during the warm-up phase was explored). The model we use
describes the physical and chemical evolution of the star-forming
region during the long pre-stellar collapse and the consequences of
the abrupt warming after the star is formed. This two-phase model is
described in detail in Sect. \ref{sec:gene_desc}.

We select a range of physical parameters intended to cover the range
likely to be appropriate for external galaxies. The metallicity and
the dust:gas mass ratio may differ substantially from the values of
the Milky Way depending on galactic mass and starburst activity. The
cosmic ray ionization rate is poorly constrained, though possibly
related to the formation rate of massive stars, and so may be large
in active galaxies. Therefore, we must choose a wide range for all
the relevant parameters; our choice is described in Sect.
\ref{sec:para}. Hot cores in the Milky Way Galaxy are remarkable for
the very high visual extinctions (typically hundreds of visual
magnitudes) associated with them. In galaxies in which the dust:gas
mass ratio is much smaller than in the Milky Way, the extinction
associated with hot cores will be smaller. However, a minimum
extinction is required for a hot core to exist, so this provides a
constraint on the choices of density distribution and physical size
of a model hot core.

Our computations describe the evolution of hot core chemistry, and we
use them to explore the sensitivity of the molecular relative (to the
total number of hydrogen atoms) abundances to the physical parameters
adopted. Our results are presented in Sect. ~\ref{sec:ana}. Trends of
the variation of hot core chemistry with changing physical parameters
are identified in that Section, and these are the main results of
this paper. In Sect. ~\ref{sec:gal} we adopt a set of values of
physical parameters that may be appropriate for several galaxy types,
and we predict hot core tracers for these types. Section
~\ref{sec:conc} gives a brief discussion and conclusion.

\section{Model description}\label{sec:gene_desc}

Hot core models have been developed by various authors to describe
the evolution of chemical abundances of various species within a
collapsing cloud during star formation. We use the hot core model
developed at University College London (hereafter called the UCL hot
core model; see \citealt{Viti99, Viti04}). To summarize, this model
is divided into two phases; the first describes the modified
free-fall cloud collapse during which time-dependent gas phase
chemistry occurs and is modified by the gas-grain interaction; it
largely follows the formulation of \citet{Rawl92}. The second phase
describes the chemical evolution after the massive star switched on.
The UCL hot core model contains a number of free parameters, and our
choices are described in detail in Sect. ~\ref{sec:para}. For these
massive stars, the switch-over from phase 1 to phase 2 of the UCL hot
core model is represented by an abrupt increase of the temperature
(for gas and grains) from 10 K (in the collapse phase) to 300 K (in
the hot core phase). In phase 2, grains and gas are assumed to be
similarly heated and the icy mantles are evaporated. Chemical
evolution in the gas phase continues, although the dynamical
evolution is assumed to have ceased. In both phases of the UCL hot
core model, the chemical network, is based on more than 1700 chemical
reactions taken from the UMIST data base \citep{Mill97, LeTe00}
involving 176 species of which 42 are surface species. Obviously, all
the molecules so far detected in galactic hot cores (e.g.
\citealt{Blac87,Mill98,Hatc98a,Font07}) have been included in this
chemistry network together with all the important and familiar
molecular species that are known from many other studies to play a
significant role in the chemistry of interstellar gas. The relevant
surface reactions included in this model are assumed to be only
hydrogenation, allowing chemical saturation where this is possible.
Note that the hot core chemistry necessarily includes species such as
CO (not characteristic of hot cores), and molecules such as H$_{2}$O
and NH$_{3}$ which are present in hot cores but undetectable.
However, these hydrides do become detectable when deuteration to HDO
and NH$_{2}$D has occurred. We have not included deuteration (nor any
other isotopologues) in the chemistry, in order to limit the size of
the chemical network. Therefore, in the discussion of results we
shall exclude water and ammonia, even though it seems clear that
these deuterated molecules should be detectable. Deuterated species
will be considered in later paper. Hydrogenation of carbon monoxide
on surfaces to methanol is also included. The efficiency we have
adopted for this reaction is chosen so that, for Milky Way
parameters, the observed fraction of CH$_{3}$OH is obtained. The rate
coefficients in current chemical network imply that methanol must be
formed in surface reactions.

Note that the results reported here are molecular abundances rather
than emission intensities. The radiative transfer calculations to
estimate intensities would require more detailed models of hot cores
and their populations than is included in our model, and the
intensities would be significantly model dependent. We do not feel
that such computations are justified in this exploratory study.

\section{Parameter choices}\label{sec:para}

The parameter choices to be made in running the hot core model are
listed in Table ~\ref{tab:2} together with the values assigned to
these parameters when representing typical hot cores in the Milky Way
(see Table ~\ref{tab:1}). We have explored the response of hot
chemistry mainly to changes in metallicity (with and without
consequent changes in the dust:gas mass ratio and H$_{2}$ formation
rate coefficient), in cosmic ray ionization rate, in hot core
temperature, in elemental relative abundances, and - since
sulfur-bearing species can be important tracers in Milky Way hot
cores - in independent variations in the sulfur abundances. We have
run over 60 hot core models, but present here results from 19
computations; the parameter choices made for these 19 calculations
are summarized in Table ~\ref{tab:2}. This table shows that we have
explored the consequences of reducing solar metallicity by factors of
up to 10$^{3}$, and independently of reducing the gas:dust ratio and
the H$_{2}$ formation rate coefficient by similar factors. We have
also explored the effect of replacing the solar metallicity by values
predicted by early Universe models, or of reducing the S/H ratio by a
factor of 10$^{2}$. We have considered the consequences on hot core
chemistry of raising the cosmic ray ionization rate by an order of
magnitude, and of increasing the hot core temperature from 300 K to
500 K. We have considered the consequences of some multiple
parameters changes. Where not stated, parameters values from Model 0
(representing hot cores in the Milky Way) are used. Note that for all
19 computations, $n_{i}$ (initial density), $n_{f}$ (final density),
$T_{1}$ (cloud temperature) and $I$ (UV radiation field intensity)
adopt the Model 0 values shown in Table ~\ref{tab:1}. There is one
proviso to note about these parameter choices. For a hot core to
form, the visual extinction associated with it must be sufficiently
optically thick at far-UV wavelengths, so that stellar UV is trapped,
warms the core and causes the ices formed during the collapse phase
to desorb. For Model 13 (which assumes a very low dust:gas mass
ratio) it is unclear as to whether this criterion is met. If the
physical dimension of the hot core is comparable to those in the
Milky Way ($\sim$ 0.03 pc) then the visual extinction for Model 13
would be about 2.6 mag, equivalent to about 8 mag in the far-UV if
the dust grains have similar optical properties to those in the Milky
Way.

\section{Sensitivity of chemical abundances to variations in the
physical and chemical parameters}\label{sec:ana}

\begin{figure*}[ht]
    \begin{center}
    \includegraphics[width=16 cm]{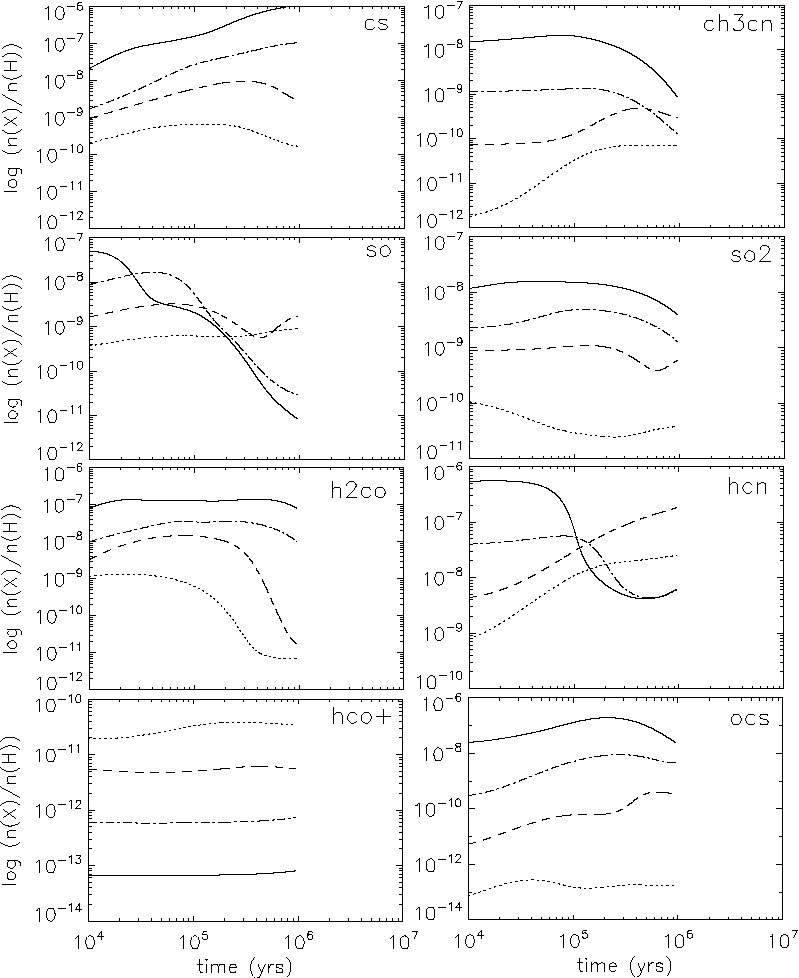}
    \caption{Influence of the metallicity alone on the logarithm of the
    relative (to the total number of hydrogen atoms n(H)) abundances (log
    [n(X)/n(H)]) of
    various species, with respect to the time (in yrs, expressed in
    logarithm scale as well). We plotted only few species which
    show the most interesting behaviors (see text in Subsect.
    ~\ref{subsec:met}). Only models from Table ~\ref{tab:2} are plotted.
    Model 0 is represented by black lines, Model 1 by dash-dot lines
    while Models 2 and 3 are symbolized by dashed lines and dotted
    lines, respectively.}\label{fig:met}
    \end{center}
\end{figure*}

\begin{figure}
     \begin{center}
     \includegraphics[width=8.7 cm]{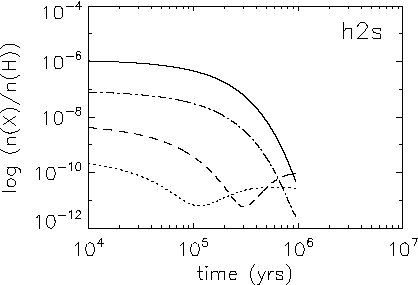}
     \caption{Influence of the metallicity alone
     on the logarithm of the relative (to the total number of
     hydrogen atoms n(H)) abundance of H$_{2}$S
     (log [n(H$_{2}$S)/n(H)]) with respect to the
     time (in yrs, expressed in logarithm scale as well). See caption
     of Fig. ~\ref{fig:met}.}\label{fig:met2}
     \end{center}
\end{figure}

The main aim of this work is to study the sensitivity of chemical
abundances in models of massive star formation to variations in
physical and chemical parameters which may be characteristic of
galaxies at different redshifts. A collection of detectable tracers
of extragalactic hot cores will be given in Subsect.
~\ref{subsec:tracer} but first we simply analyze the trends that the
chemistry follows as each parameter was varied. Figures
~\ref{fig:met} to ~\ref{fig:SH2} show examples of the most
interesting changes while Tables ~\ref{tab:5}, ~\ref{tab:5b},
~\ref{tab:6} and ~\ref{tab:7} summarize the temporal trends of
fractional abundances with variation in metallicity, in cosmic ray
ionization rate, and in initial elemental abundances. In the
following discussion, we rather arbitrarily assume that molecules
with fractional abundances less than about 10$^{-12}$ are probably
undetectable, a criterion that is roughly satisfied in the Milky Way.

\subsection{Sensitivity to variations in metallicity}\label{subsec:met}

Models where changes in metallicity alone (Models 0 to 3 in Table
~\ref{tab:2}) as well as (more realistically) changes in metallicity
coupled with other parameters (Models 11 to 13 in Table ~\ref{tab:2})
have been investigated. The first set of models were run in order to
isolate the effects due solely to metallicity changes (see Figs.
~\ref{fig:met}, ~\ref{fig:met2} and Tables ~\ref{tab:5}, ~\ref{tab:5b}).

Reducing metallicity leads to no great surprises for most species in
that their abundance is simply reduced accordingly; however, one
notes that as the metallicity is reduced to 1\% of solar the
fractional abundances of some species such as CS tend to decrease
with time (with the released atomic sulfur going into SO); while for
the solar case and 1/10th of the solar it continues to increase with
time.

However, some species do not seem as sensitive to metallicity
changes: for example, the fractional abundances of CH$_3$CN for
different metallicities tend to converge after $\sim$ 10$^5$ yrs,
making this species rather insensitive to metallicity and therefore
(potentially) a good tracer of hot cores at high redshift. This
`convergence' is due to a faster destruction (due to cosmic rays) at
high metallicities (where more ions are present). SO and H$_2$S also
tend to converge. However, SO conversion into SO$_2$ by cosmic
ray-induced reactions is inevitably slowed down at low metallicity,
so that SO may be more abundant at late times for lower metallicity
than solar.

A good discriminator of metallicity is H$_2$CO. Its abundance varies
proportionally to metallicity as metallicity is reduced from solar
values, and then drops below detectable values for metallicities less
than 1/100th of solar. Another interesting species is HCN whose
fractional abundance is directly proportional to metallicity up to
$\sim$ 10$^5$ yrs but then this behavior is reversed, i.e. its
abundance becomes larger at lower metallicity. This is probably due
to an increase in ionization for lower metallicity, as can be seen
from the behavior of HCO$^+$. The latter, not usually a hot core
tracer, is in fact probably only detectable at very low ($\sim$ 0.1\%
of solar) metallicity.

Perhaps more significant may be the trends among models where the
metallicity variations are also applied directly to the dust:gas mass
ratio and the H$_{2}$ formation rate. However, we find that most
chemical abundances behave as in the simpler models where only
metallicity varies (see Tables ~\ref{tab:5} and ~\ref{tab:5b}). It is
worth noting that OCS and SO$_2$ abundances are fairly insensitive to
metallicity changes and are good tracers of hot cores provided the
metallicity does not drop below about 1\% of solar.

\subsection{Sensitivity to variations in the cosmic ray ionization
rate}\label{subsec:zeta}

\begin{figure*}
    \begin{center}
    \includegraphics[width=16.5 cm]{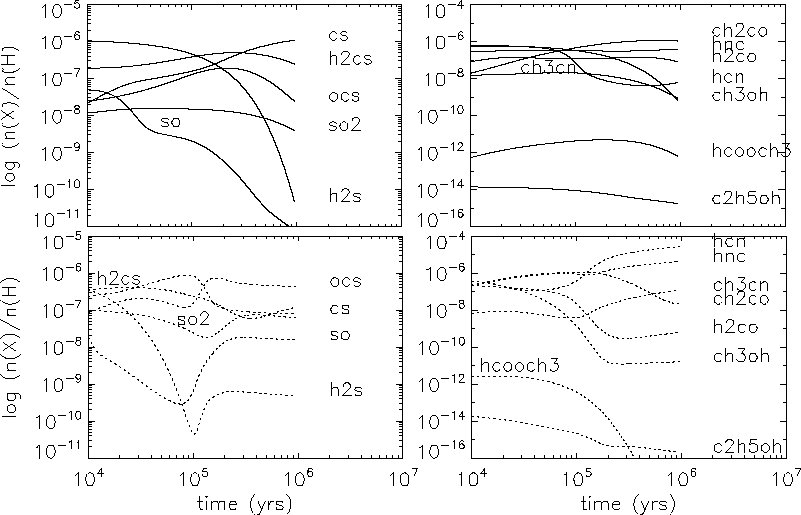}
    \caption{Influence of the cosmic ray ionization rate alone
    on the logarithm of the relative (to the total number of
    hydrogen atoms n(H)) abundances
    (log [n(X)/n(H)]) of various species
    with respect to the time (in yrs, expressed in logarithm scale
    as well). See the caption of the Fig. ~\ref{fig:met} as
    well as the text in the Subsect. ~\ref{subsec:zeta}. We
    plotted two models whose parameters
    are listed in Table ~\ref{tab:2} : Model 0 is
    represented by black lines (top) and Model
    14 ($\zeta = 10 \times \zeta_{\odot}$) corresponds to
    the dotted lines (bottom).}\label{fig:cosmic}
    \end{center}
\end{figure*}

The following results concern Models 0 and 14 (defined in Table
~\ref{tab:2}). Our results are shown in Fig ~\ref{fig:cosmic} and
summarized in Table ~\ref{tab:6}. The effects of increasing the
cosmic ray ionization rate are complex and in general lead to a
chemistry that approaches steady-state more quickly (see Fig.
~\ref{fig:cosmic}). For the sulfur-bearing species, a higher $\zeta$
implies a faster destruction of CS and hence a faster increase in the
abundances of its daughters species such as SO, SO$_2$ and OCS.
Nitrogen-bearing species such as HCN and CH$_3$CN increase with
$\zeta$ while oxygen bearing species such as CH$_3$OH, H$_2$CO and
CH$_2$CO decrease. Some species, such as water and CO, are not
significantly affected by the larger ionization rate, since
conversion to these species by both surface and gas-phase chemistries
is complete in Phase 1. It is interesting to note how oxygen and
nitrogen bearing species are anti-correlated as a function of
$\zeta$, probably due to the high abundance of ionized carbon in the
case with higher ionization rate, and this reacts quickly with many
oxygen-bearing species. The main reservoir of nitrogen is N$_{2}$
which, when ionized, gives N$^{+}$, a reactive species that promotes
the formation of nitrogen-bearing molecules.

\subsection{Sensitivity to variations in temperature}\label{subsec:temp}

\begin{figure*}
    \begin{center}
    \includegraphics[width=16.5 cm]{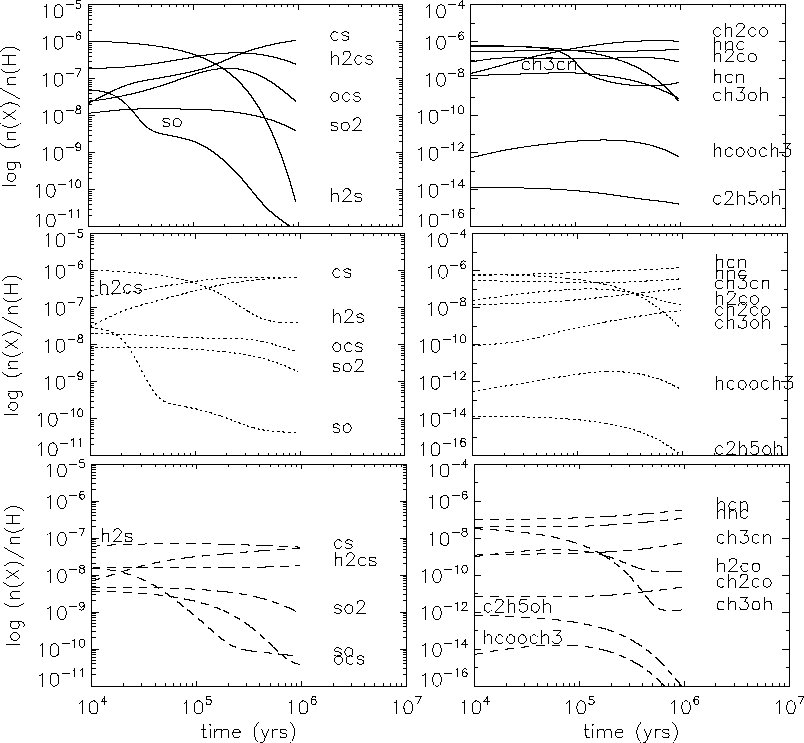}
    \caption{Influence of the hot core temperature alone on the
    logarithm of the relative (to the total number of
    hydrogen atoms n(H)) abundances
    (log [n(X)/n(H)]) of various species with
    respect to the time (in yrs, expressed in logarithm scale as well).
    See the caption of the Fig. ~\ref{fig:met} as
    well as the text in the Subsect. ~\ref{subsec:temp}. We plotted
    three models whose parameters are listed in
    Table ~\ref{tab:2} : Model 0 is represented
    by black lines (top two plots) while Model 15 corresponds to dotted
    lines (middle two plots). Model 0 and 15 only differ in the
    value of their hot core temperature. The bottom
    two plots are for Model 16 (dashed lines, low-metallicity
    T$_{2}$=500 K, see Table ~\ref{tab:2}).}\label{fig:temp}
    \end{center}
\end{figure*}

We ran two models in which the hot core temperature, T$_{2}$, was
increased from 300 K to 500 K (Models 15, with solar metallicity, and
Model 16 with 1/10th of solar metallicity). In order to isolate the
temperature variations effects, we first compare two solar
metallicity models where only the temperature is varied (see the top
two plots of the Fig. ~\ref{fig:temp}). As expected, several species
are affected by an increase in hot core temperature: H$_2$S, HCN and
CH$_3$CN all survive for much longer making them good tracers of
temperature. CH$_2$CO abundance, on the other hand, is 2 orders of
magnitude lower in the high temperature case: this is due to its
destruction by cosmic ray induced photo-dissociation which has a
significant temperature dependence.

At lower metallicity (see the bottom plot of the Fig.
~\ref{fig:temp}), it is interesting to note that H$_2$S remains high
despite the lower abundance of metals and in fact it is the most
abundant sulfur-bearing species up until 10$^6$ years when CS becomes
comparable: hence, for a higher temperature and low metallicity hot
core H$_2$S remains a good tracer.

\subsection{Sensitivity to variations in initial elemental
abundances}\label{subsec:abund}

\begin{figure*}
    \begin{center}
    \includegraphics[width=16 cm]{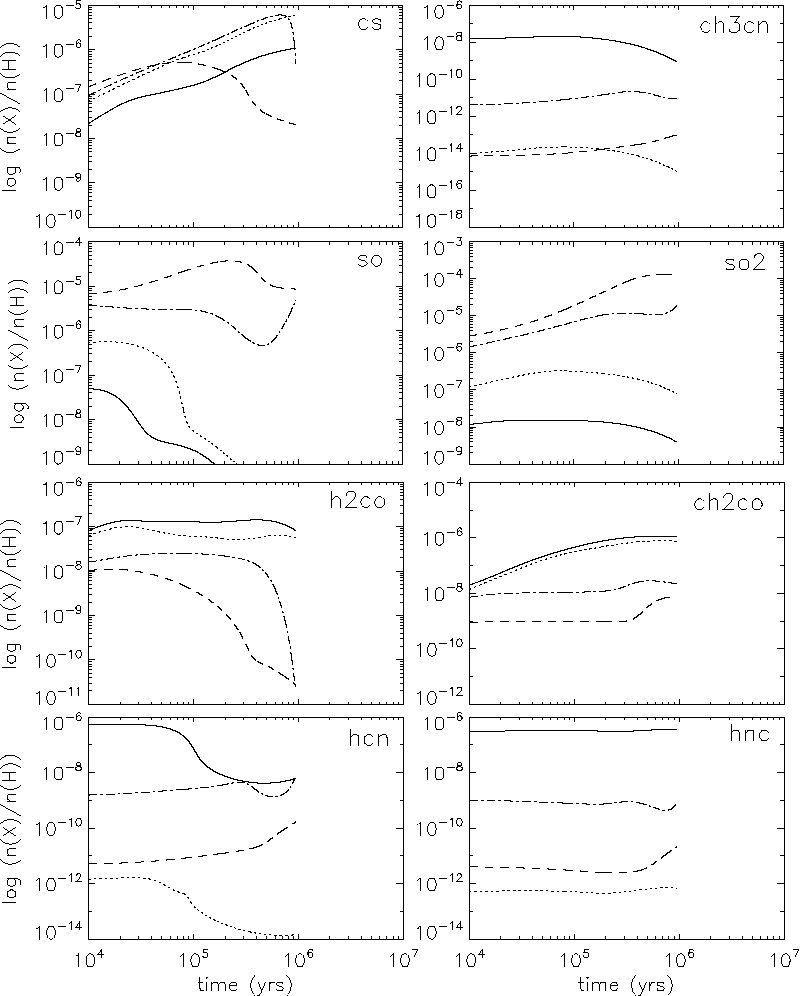}
    \caption{Influence of the initial elemental abundances ratios alone on the
    logarithm of the relative (to the total number of hydrogen atoms n(H))
    abundances (log [n(X)/n(H)]) of various species
    with respect to the time (in yrs, expressed in logarithm
    scale as well). See the caption of the Fig. ~\ref{fig:met} as
    well as the text in the Subsect. ~\ref{subsec:abund}. We plotted
    four models whose parameters are listed in Table \ref{tab:2}
    : Model 0 which is represented by black lines,
    Model 4 which is symbolized with dotted lines and the
    Models 5 and 6 which correspond to dashed lines and
    dash-dot lines, respectively.}\label{fig:abundratio}
    \end{center}
\end{figure*}

\begin{figure*}
    \begin{center}
    \includegraphics[width=16 cm]{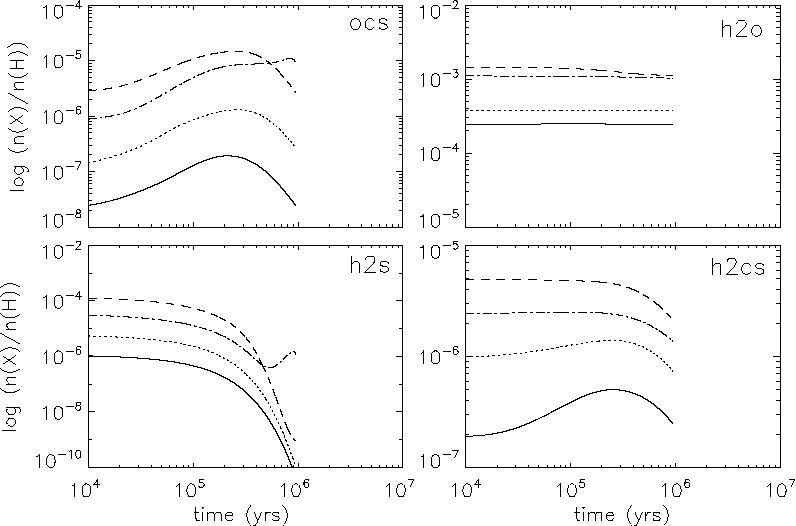}
    \caption{Influence of the initial elemental abundances
    ratios alone on the logarithm of the relative (to the total number
    of hydrogen atoms n(H)) abundance of various species
    (log [n(X)/n(H)]) with respect to the time
    (in yrs, expressed logarithm scale as well). See caption of
    Fig. ~\ref{fig:abundratio}.}\label{fig:abundratio2}
    \end{center}
\end{figure*}

\begin{figure*}
    \begin{center}
    \includegraphics[width=16 cm]{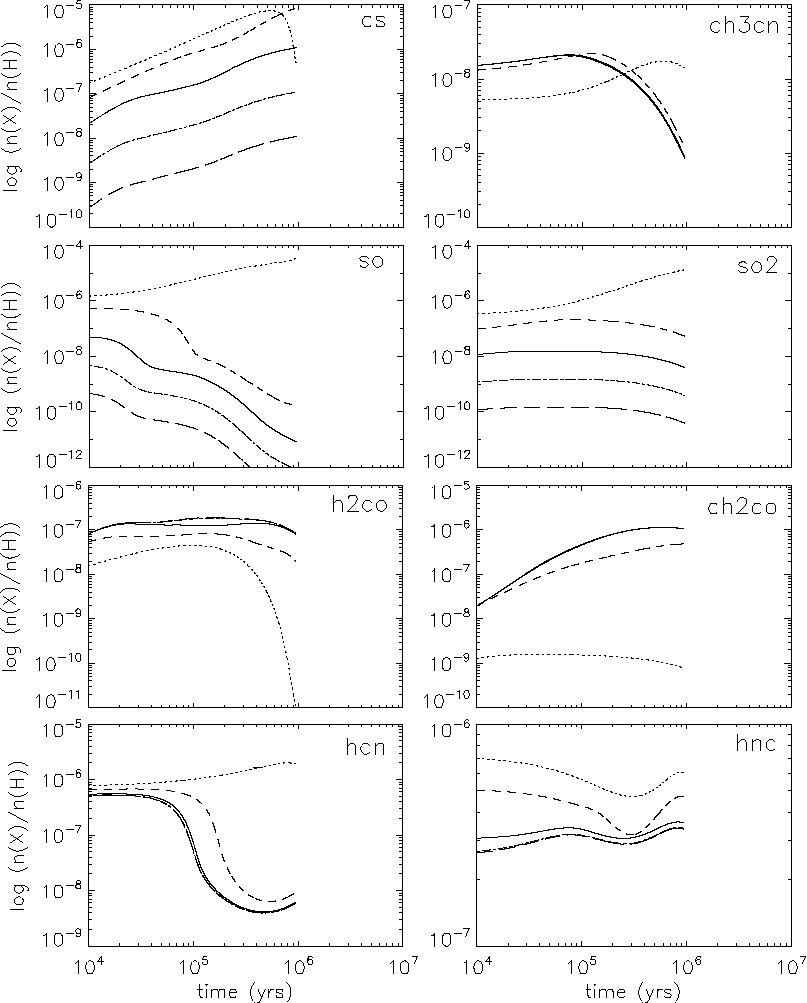}
    \caption{Influence of the initial elemental abundance
    S/H ratio alone on the
    logarithm of the relative (to the total number of hydrogen atoms n(H)) abundances
    (log [n(X)/n(H)]) of various species with respect
    to the time (in yrs, expressed in logarithm scale as well). See the
    caption of the Fig. ~\ref{fig:met} as well as the text in the
    Subsect. ~\ref{subsec:abund}. We plotted four models whose parameters
    are listed in Table ~\ref{tab:2}: Model 0 which is represented by black lines,
    Model 7 which is symbolized with dotted lines and the Models 8, 9 and
    10 which correspond to small dashed lines, dash-dot lines and
    large dashed lines, respectively.}\label{fig:SH}
    \end{center}
\end{figure*}

\begin{figure*}
    \begin{center}
    \includegraphics[width=16 cm]{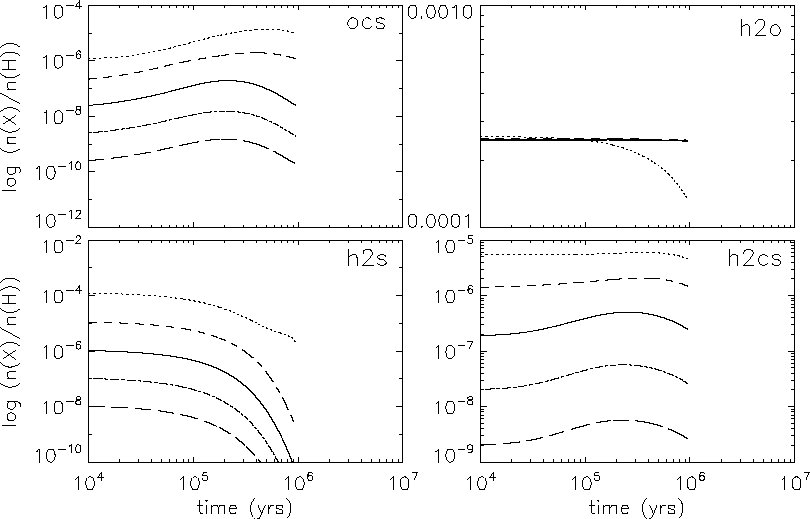}
    \caption{Influence of the initial elemental abundance
    S/H ratio alone on the logarithm of the relative
    (to the total number of hydrogen atoms n(H)) abundance various species
    with respect to the time (in yrs, expressed
    logarithm scale as well). See caption of
    Fig. ~\ref{fig:SH}.}\label{fig:SH2}
    \end{center}
\end{figure*}

The initial elemental abundances (i.e. before star formation occurs)
of galaxies at high redshifts are by no means well-known. We
therefore consider briefly the sensitivity of molecular abundances in
the hot core phase to variations in the initial abundance of the main
elements (see Models 4 to 10 in the Table ~\ref{tab:2}). We consider
the consequences of adopting the elemental abundances arising from
stellar evolution models in initially zero metallicity gas, as
proposed by \citet{Chie02, Hege02, Umed02} (see Table ~\ref{tab:3}).
We have also investigated the effect of varying independently the
initial sulfur abundance. The results for this section are shown in
Figs. ~\ref{fig:abundratio}, ~\ref{fig:abundratio2},
~\ref{fig:SH} and ~\ref{fig:SH2} and summarized in Table ~\ref{tab:7}.

Beside the expected changes in absolute abundances (e.g. H$_2$O
abundance increases for oxygen-rich models), the most striking
variations are, not surprisingly, shown by sulfur-bearing species, in
particular (see Figs. ~\ref{fig:abundratio} and
~\ref{fig:abundratio2}) : while for most models CS remains abundant
for the (canonical) lifetime of the core, in the HW02 model it
decreases steeply (with atomic sulfur going into SO and SO$_{2}$)
after 10$^{4}$ yrs. This suggests that in an oxygen-rich and
sulfur-rich environment CS would not be a good tracer of massive
star-forming regions while SO, and in particular, SO$_2$ would be
enhanced with respect to galactic hot cores. The chemical evolution
of OCS follows the same trend to late times ($\sim$ 10$^{6}$ yrs) for
all initial abundances, but its abundance is directly proportional to
the amount of sulfur initially available so it could be a good
indicator of the initial cosmic sulfur abundance. The same behavior
is found for CH$_3$CN which follows the nitrogen abundance so, if the
nitrogen-deficient models of Table ~\ref{tab:3} are correct, this
molecule is unlikely to be detectable in high redshift hot cores. The
sulfur bearing species SO and H$_2$S actually increase in abundance
after $\sim$ 10$^6$ yrs for Model 6. This is an oxygen-rich model
where sulfur is close to solar. Finally, although largely insensitive
to variations of the initial elemental abundances, H$_2$CS is
nevertheless always quite abundant and may therefore be used as a
tracer of hot cores regardless of the chemical composition of the
galaxy in question.

Of other species, H$_2$CO, HCN and HNC are of particular interest.
The H$_2$CO abundance is fairly flat in its time evolution for
oxygen-rich models, at least up to 10$^{6}$yrs, unless the oxygen
abundance is very high (as in Model 15). In that case, the H$_2$CO
abundance falls steadily between 10$^{5}$yrs and 10$^{6}$yrs. HCN
decreases with time for oxygen-rich models, while it increases for
oxygen-poor models, as HCN is directly destroyed by atomic oxygen.

In galactic star-forming regions, the sulfur initial elemental
abundance is a particularly poorly-determined parameter
\citep{Ruff99, Wake04} and yet sulfur-bearing species are often used
as 'chemical clocks'. The main conclusion of varying solely the
initial sulfur abundance by 4 orders of magnitude below solar (Models
7 to 10 in the Table ~\ref{tab:2}; the results are shown in Figs.
~\ref{fig:SH} and ~\ref{fig:SH2}, with summary comments in Table
~\ref{tab:7}) is that non sulfur-bearing species are only sensitive
to the initial sulfur abundance if the latter is comparable to
carbon. So, for example, CH$_3$CN and HCN do not decline as fast when
the initial abundances of carbon and sulfur are comparable, remaining
abundant enough to be detected even at late times, while the opposite
is true for H$_2$CO which declines after 10$^5$ yrs and for CH$_2$CO
which never achieves high abundances in Model 0. Sulfur-bearing
species are of course heavily affected by the initial abundance of
sulfur in a fairly predictable way, but it is worth noticing that SO
and SO$_2$ increase with time for the solar case (Model 0) while they
decrease in every other model. This may imply that, although commonly
considered as robust "chemical clocks", sulfur-bearing species may
only be useful in this respect for star-forming regions closely
resembling those in the Milky Way Galaxy.

\subsection{Tracers of extragalactic hot
cores}\label{subsec:tracer}

Tables ~\ref{tab:5}, ~\ref{tab:5b}, ~\ref{tab:6} and ~\ref{tab:7}
show that most of
the molecules familiar from studies of hot cores in the Milky Way
Galaxy are also likely to have significant abundances in hot cores in
external active galaxies, even though the physical and chemical
conditions in those galaxies may be very different from those in the
Milky Way. We have identified in those tables molecules that should
trace in hot cores in external galaxies the local metallicity, cosmic
ray ionization rate, hot core temperature and relative elemental
abundances. The molecules are CS, OCS, CH$_3$CN, CH$_3$OH, SO,
SO$_{2}$, H$_{2}$S, H$_{2}$CS, H$_{2}$CO, CH$_{2}$CO, HCN and HNC.
Ethanol and methyl formate are also known as hot core molecules in
the Milky Way Galaxy. However, according to our hot core models, they
appear to be generally of rather low abundance in external galaxies.
It is however possible that our models do not account adequately for
the formation (by surface chemistry) of these two species under
conditions appropriate for external galaxies.

For galaxies with metallicities lower than or comparable to that of
the Milky Way, the molecules with highest abundances (and therefore
likely to be useful hot core tracers) include CS, SO$_{2}$,
CH$_{2}$CO and HNC. For metallicities that are at least 1\% of that
of the Milky Way, the following molecules are also be likely to be
useful hot core tracers : OCS, H$_{2}$CS and H$_{2}$CO. The molecule
SO can also be abundant, but tends to be an intermediate species in
the formation of SO$_{2}$ during the hot core lifetime. We have
already noted the convergence of the abundance of CH$_3$CN to a value
that is rather insensitive to the metallicity; this value is rather
low ($\sim$ 10$^{-10}$) but still significant.

For galaxies with enhanced cosmic ray fluxes but otherwise similar
parameters to those of the Milky Way, the most abundant species are
likely to be OCS, H$_{2}$CS, CS and SO$_{2}$; these molecules between
them contain most of the available sulfur. Likewise, HCN, HNC and
CH$_3$CN contain large amounts of the available nitrogen. CH$_{2}$CO
and H$_{2}$CO have large abundances that are enhanced when the cosmic
ray flux is raised.

When the temperature of the hot cores is raised, the ratio of the
abundances of HCN and HNC approaches unity.

It is interesting to note that even when the initial elemental
abundances are those arising from the first generation of stars, hot
core molecular abundances can be high. Model 4 shows that both
CH$_{2}$CO and H$_{2}$CO should be prominent tracers.

\section{Selected galaxies and predicted hot core tracers}\label{sec:gal}

In this section, we select several fairly well observed galaxies,
representative of different morphological types, that may be
reasonably compared with some of the model galaxies listed in Table
~\ref{tab:2}. Our aim is to identify the predicted tracers of hot
cores in these galaxies.

We identify Model 0 as a potentially good representative normal
spiral. The metallicity of IC 342 is estimated to be 1.07 z$_{\odot}$
(using 12+log(O/H) = 9.30 from \citealt{Vila92, Garn98b}) while for
NGC 4736 we have 1.04 z$_{\odot}$ (using 12+log(O/H) = 9.01 from
\citealt{Zari94}). In addition, these two galaxies do not show any
bursts of star formation, suggesting that a radiation field
comparable to that of the Milky Way as well as a hot core temperature
of 300 K may be appropriate for them. They do not contain any
additional source of energy such as, for instance, an AGN which could
be responsible to an enhancement of the cosmic ray ionization rate
value. All these physical parameter values suggest that Model 0 may
be a good representative for these two galaxies.

A second interesting case is Model 15 having solar metallicity, a
standard radiation field, hot core temperature of 500 K and normal
cosmic ray ionization rate. It could be an appropriate model for
starburst galaxies such as M 83 (metallicity of 1.06 z$_{\odot}$
derived from value in \citealt{Zari94}) since starburst environments
are expected to show higher gas temperatures without necessarily any
additional sources of energy justifying an enhancement of the cosmic
ray ionization rate (except in the presence of an AGN which is
suggested to be a potential source of high energy cosmic rays).
Taking a standard value for the radiation field for this kind of
source may not be appropriate but, as the metallicity is solar, the
gas to dust mass ratio is 100, the final H$_{2}$ number density is
1$\times$10$^{7}$ cm$^{-3}$, the visual extinction associated with a
hot core is $\approx$ 500-600 mag. Therefore, the radiation field has
negligible influence on hot core tracers in such galaxy types.

Galaxies of the type of IC10 are also interesting to study.
\citet{Garn98b} found a value of IC 10 metallicity $\approx
\frac{1}{5}$ z$_{\odot}$. This source is usually considered as one of
the closest starburst sources in the Universe even if IC 10 has been
recently recognized as only marginally starburst \citep{Hida05}.
Adding the presence of numerous Wolf-Rayet stars in IC 10, we could
however suppose a temperature higher than in quiet galaxy.
Consequently, a model with a z=$\frac{1}{5}$ z$_{\odot}$, temperature
of hot core of 500 K, $\zeta$ = 1 (no AGN) appears appropriate. As in
the previous case, the radiation field is largely irrelevant to the
hot core chemistry. We suggest IC10 could be represented by Model 17.

It would be interesting to identify a model for high redshift sources
such as the Cloverleaf (redshift of z$_{red.} \approx$ 2.6) or APM
08279 (z$_{red.} \approx$ 3.9). For these sources, the physical
properties are unknown. However, molecular detections already
obtained \citep{Gao04b, Wu05, Wagg05} suggest that star formation may
be active, even if these galaxies are affected by their AGNs.
\citet{Garc06} have suggested that the unusual chemistry they have
been observed may be due to a combination of star formation and AGN
(PDR/XDR) chemistries. In an attempt to describe such sources, we
adopt in Model 18 a high value of the hot core temperature, a high
cosmic ray ionization rate, a high value of the radiation field
intensity, and a relatively low metallicity (all with respect to the
Milky Way).

Table ~\ref{tab:8} summarizes the typical molecular tracers of
extragalactic hot cores which are detectable for these four kinds of
sources, where the detectability limit is arbitrarily taken to be
equivalent to a fractional abundance of 1$\times 10^{-12}$. Here our
goal was to give to the reader some observational clues derived from
this modelling work. The table indicates that these should be some
differences in the appropriate tracers to use for each galaxy type.
The overall trends (described in Sect. ~\ref{sec:ana}) of those
tracers with physical parameters should enable a more detailed
description of the galaxy to be made.

In the two lowest metallicity cases represented in Table ~\ref{tab:8}
(i.e. Models 17 and 18) some of the relative abundances of potential
tracers are surprisingly large. For example, Model 17 predicts that
in IC10 CS, H$_{2}$S, HCN and HNC should all have fractional
abundances $\sim$ 10$^{-8}$, while CH$_3$CN is about one order of
magnitude smaller. Model 18 predicts that HCN and HNC should have
fractional abundances $\sim$ 10$^{-7}$, CS $\sim$ 10$^{-8}$, and
H$_{2}$CS and CH$_3$CN both $\sim$ 10$^{-9}$. According to the
estimates of \citet{Lint05} these abundances are sufficiently large
that unresolved active galaxies should be detectable in these species
even at high redshift. Observational tests of prediction made in
Table ~\ref{tab:8} will be described in a forthcoming paper. Recent
observations of Arp220 \citep{Aalt02, Aalt07} find large abundances
of some hot core tracers such as HC$_{3}$N. It is yet not clear
whether the chemistry is dominated by UV, X-rays or other physical
processes. Model 1 or Model 16 may best represent this galaxy.

Gravitational lensing of star-forming regions in high z galaxies may
significantly affect the molecular line intensities observed.
However, relative line intensities should be preserved, and the hot
core `signature' should still be discernable.

\section{Conclusion}\label{sec:conc}

We have computed the chemical abundances in hot cores with a wide
range of physical parameters. In particular, we have discussed the
sensitivity of molecular abundances to changes in metallicity, to the
local cosmic ray ionization rate, to hot core temperature, and to the
relative elemental abundances that may represent the cosmic
composition within galaxies early in the evolution of the Universe.
In all cases that we have examined there is a rich hot core
chemistry, even when the metallicity is low. The fractional
abundances of our predicted tracers are at a level that was shown by
\citet{Lint05} to provide for an active galaxy (containing many hot
cores) - even at high redshift - a detectable signal, though of
course unresolved.

The chemical network within a hot core is complex, and not all
species respond in the same way to changes in physical parameters. We
find, for example, that some species trace linearly any changes in
metallicity, while others may respond inversely to such changes. Many
species are enhanced in abundance if the flux of cosmic rays is
increased but other species are reduced in abundance.

We have created models of hot core chemistry that describe crudely
several reasonably well-observed galaxies. We have predicted hot core
tracers for these galaxy types. Detection of hot core species should
- with the use of models such as those presented here - provide
detailed descriptions of the physical conditions within active
galaxies.

\begin{table*}
    \caption{Standard model parameters (Model 0; see Sect. ~\ref{sec:para}).}\label{tab:1}
    \begin{center}
    \begin{tabular}{c c c c c}
    \hline
    Parameter & Symbol & Typical Milky\\
    & & Way values\\
    \hline
    Free fall modifier & $B$ & 0.1\\
    Initial number density (phase 1)& $n_{i}$ & 300 H cm$^{-3}$\\
    Final number density (phase 1 and 2)& $n_{f}$ & 1$\times 10^{7}$ H cm$^{-3}$\\
    Temperature (phase 1) & $T_{1}$ & 10 K \\
    Temperature (phase 2) & $T_{2}$ & 300 K\\
    External UV radiation intensity & $I$ & 1 Habing \\
    Cosmic ray ionization rate & $\zeta$ & 1.3$\times 10^{-17}$
    s$^{-1}$\\
    Visual extinction & $A_{v}$ & 600 mag\\
    Gas:dust ratio & $d$ & 100 \\
    H$_{2}$ formation rate coefficient & $R$ & 1.0$\times 10^{-17}\times \sqrt{T}$ cm$^{3}$s$^{-1}$\\
    Metallicity & z$_{\odot}$ & solar values, see Table ~\ref{tab:2}\\
    \hline
    \end{tabular}
    \end{center}
\end{table*}

\begin{table*}
    \caption{Input parameters of the UCL hot core models (see
    Sect. ~\ref{sec:para} and the text and the figures in Sect. ~\ref{sec:ana}).
    These
    19 models have been selected because they are the most relevant
    models for studying the influence of the input parameters of the UCL
    hot core model on the predicted relative (to the total number of
    hydrogen atoms n(H)) abundances
    (n(X)/n(H)) of various species. The
    abbreviation ``ST'' represents the standard values listed in
    Table ~\ref{tab:1} while the abbreviations ``CL02, HW02 and UN02''
    are elemental abundances ratios references presented in
    Table ~\ref{tab:3}. The abbreviation ``CB-A, -B, -C, -D and -E''
    correspond to the values listed in Table ~\ref{tab:4}.}\label{tab:2}
    \begin{center}
    \begin{tabular}{c c c c c c c c}
    \hline
    Model & Metallicity$^{a}$ & Gas-to-dust & Ini. Elem. & S/H & $\zeta ^{b}$ & Temp. &
    Comb.\\
    & (z$_{\odot}$) & mass ratio & Abund. ratios && ($\zeta_{\odot}$) & hot core (K) &
    parameters\\
    \hline
    0 & 1 & 100 & ST & 1.4$\times$10$^{-6}$ & 1 & 300 & ST\\
    1 & 1/10 & 100 & ST/10 & 1.4$\times$10$^{-7}$ & 1 & 300& ST\\
    2 & 1/100 & 100 & ST/100 & 1.4$\times$10$^{-8}$ & 1 & 300& ST\\
    3 & 1/1000 & 100 & ST/1000 & 1.4$\times$10$^{-9}$ & 1 & 300& ST\\
    4 & 1 & 100 & CL02 & 7.59$\times$10$^{-6}$ & 1 & 300& ST\\
    5 & 1 & 100 & HW02 & 1.54$\times$10$^{-4}$ & 1 & 300& ST\\
    6 & 1 & 100 & UN02 & 4.06$\times$10$^{-5}$ & 1 & 300& ST\\
    7 & 1 & 100 & ST & 1.4$\times$10$^{-4}$ & 1 & 300& ST\\
    8 & 1 & 100 & ST & 1.4$\times$10$^{-5}$ & 1 & 300& ST\\
    9 & 1 & 100 & ST & 1.4$\times$10$^{-7}$ & 1 & 300& ST\\
    10 & 1 & 100 & ST & 1.4$\times$10$^{-8}$ & 1 & 300& ST\\
    11 & 1/10 & 1000 & ST/10 & 1.4$\times$10$^{-7}$ & 1 & 300& CB-A\\
    12 & 1/100 & 1$\times$10$^{4}$ & ST/100 & 1.4$\times$10$^{-8}$ & 1 & 300& CB-B\\
    13 & 1/1000 & 1$\times$10$^{5}$ & ST/1000 & 1.4$\times$10$^{-9}$ & 1 & 300& CB-C\\
    14 & 1 & 100 & ST & 1.4$\times$10$^{-6}$ & 10 & 300 & ST\\
    15 & 1 & 100 & ST & 1.4$\times$10$^{-6}$ & 1 & 500 & ST\\
    16 & 1/10 & 1000 & ST/10 & 1.4$\times$10$^{-7}$ & 1 & 500 & CB-A\\
    17 & 1/5 & 500 & ST/5 & 2.8$\times$10$^{-7}$ & 1 & 500 & CB-D\\
    18 & 1/25 & 2500 & ST/25 & 5.6$\times$10$^{-8}$ & 10 & 500 & CB-E\\
    \hline
    \end{tabular}
    \end{center}
    $^{a}$ : z$_{\odot}$ = 1 corresponds to solar values of the elemental
    abundances ratios; $^{b}$ : expressed in units of
    $\zeta_{\odot}$ = 1.3$\times 10^{-17}$ s$^{-1}$.
\end{table*}

\begin{table}
    \caption{Initial abundance ratios values used in Table ~\ref{tab:2}.
    The abbreviations ``CL02, HW02 and UN02'' refer to \citet{Chie02}, \citet{Hege02}
    and \citet{Umed02}, respectively (see Sect. ~\ref{sec:para}).
    The standard initial abundance ratios values are from
    \citet{Sava96, Sofi97, Meye98, Snow02, Knau03}. We did not include in
    this table the values of the initial elemental abundance S/H
    ratio since they have been already listed in Table \ref{tab:2}.
    }\label{tab:3}
    \begin{center}
    \begin{tabular}{c c c c c}
    \hline
    & ST & CL02 & HW02 & UN02 \\
    \hline
    C/H & 1.4$\times$10$^{-4}$ & 1.4$\times$10$^{-4}$ & 1.4$\times$10$^{-4}$ &
    1.4$\times$10$^{-4}$\\
    O/H & 3.2$\times$10$^{-4}$ & 4.54$\times$10$^{-4}$ & 1.53$\times$10$^{-3}$ &
    1.18$\times$10$^{-3}$\\
    N/H & 6.5$\times$10$^{-5}$ & 5.99$\times$10$^{-11}$ & 1.58$\times$10$^{-9}$ &
    3.24$\times$10$^{-7}$\\
    He/H & 7.5$\times$10$^{-2}$ & 7.5$\times$10$^{-2}$ & 7.5$\times$10$^{-2}$ &
    7.5$\times$10$^{-2}$\\
    Mg/H & 5.1$\times$10$^{-6}$ & 1.83$\times$10$^{-5}$ & 1.2$\times$10$^{-4}$ &
    6.18$\times$10$^{-5}$\\
    \hline
    \end{tabular}
    \end{center}
\end{table}

\begin{table*}
    \caption{Combinations of parameters coupled with the metallicity and used in the
    Table. ~\ref{tab:2} (see Sect. ~\ref{sec:para}).}\label{tab:4}
    \begin{center}
    \begin{tabular}{c c c c c c c}
    \hline
    & ST & CB-A & CB-B & CB-C & CB-D & CB-E\\
    \hline
    $A_{v}$ & 581.1 & 59.9 & 7.8 & 2.6 & 117.8 & 25.2\\
    (mag) &&&&\\
    \hline
     ratio of the number &&&&&&\\
     densities of grains & 1.0$\times$10$^{-12}$ & 1.0$\times$10$^{-13}$ &
    1.0$\times$10$^{-14}$ & 1.0$\times$10$^{-15}$ & 2.0$\times$10$^{-13}$
    & 4.0$\times$10$^{-14}$ \\
    to hydrogen nuclei&&&&&&\\
    \hline
    H$_{2}$ form.$^{a}$ & 1.73$\times$10$^{-16}$ & 1.73$\times$10$^{-17}$ & 1.73$\times$10$^{-18}$ &
    1.73$\times$10$^{-19}$ & 6.32$\times$10$^{-18}$ & 1.26$\times$10$^{-18}$\\
    rate coeff. &&&&\\
    \hline
    \end{tabular}
    \end{center}
    $^{a}$ : The H$_{2}$ formation rate coefficient is computed at the beginning
    of the phase 2 of the UCL hot core model and it is
    expressed in cm$^{3}$s$^{-1}$.
\end{table*}

\begin{table*}
    \caption{Temporal trends of molecular fractional abundances,
    computed for different values of metallicity. Other model parameters have the standard
    values listed in Table ~\ref{tab:1}.}\label{tab:5}
    \begin{center}
    \begin{tabular}{l l}
    \hline
    Molecule & Response to metallicity changes\\
    \hline
    CO, H$_2$O & good tracers of metallicity since flat \\
    CS & good tracer of metallicity since grows strongly during phase
    2\\
    OCS & undetectable at lowest metallicity, otherwise, a good
    metallicity tracer\\
    CH$_3$CN & tends to converge at late times in hot core epoch; \\
    & roughly independent of the metallicity\\
    CH$_3$OH & good tracer of metallicity, but lost at late times\\
    SO & converted at late time to SO$_2$, most rapidly for high
    metallicity cases. \\
    & An inverse tracer of metallicity at late hot core epoch\\
    SO$_2$ & detectable at all metallicities; a linear metallicity
    tracer\\
    H$_2$S & at early times a linear metallicity tracer; lost at late times \\
    & by chemical conversion\\
    H$_2$CS & a strong indicator of metallicity, $\approx$ (metallicity)$^{2}$\\
    H$_2$CO & a linear indicator of metallicity\\
    CH$_2$CO & a linear indicator of metallicity\\
    C$_2$H$_5$OH & undetectable\\
    HCOOCH$_3$ & undetectable\\
    HNC & abundant, but poor metallicity tracer\\
    HCN & abundant, but poor metallicity tracer; chemical conversion \\
    & at late times\\
    HCO$^+$ & good inverse tracer of metallicity, highest for lowest metallicity; \\
    & undetectable for solar metallicity\\
    \hline
    \end{tabular}
    \end{center}
\end{table*}

\begin{table*}
    \caption{Temporal trends of molecular fractional abundances, computed for different values of the metallicity coupled with other parameters (see Table ~\ref{tab:4}).}\label{tab:5b}
    \begin{center}
    \begin{tabular}{l l}
    \hline
    Molecule & Response to changes in metallicity, dust:gas mass ratio,\\
    & and H$_{2}$ formation rate coefficient\\
    \hline
    CO, H$_2$O & trends as in Table ~\ref{tab:5}\\
    CS & trends as in Table ~\ref{tab:5}\\
    OCS & undetectable at lowest metallicity, otherwise, less sensitive \\
    & to metallicity than in Table ~\ref{tab:5}\\
    CH$_3$CN & fairly independent of metallicity\\
    CH$_3$OH & traces metallicity; declines steeply after 10$^{5}$yrs\\
    SO & trends as in Table ~\ref{tab:5}\\
    SO$_2$ & independent of metallicity for higher metallicity values\\
    H$_2$S & trends as in Table ~\ref{tab:5}\\
    H$_2$CS & a strong metallicity indicator\\
    H$_2$CO & trends as in Table ~\ref{tab:5}\\
    CH$_2$CO & follows metallicity\\
    C$_2$H$_5$OH & undetectable\\
    HCOOCH$_3$ & undetectable\\
    HNC & trends as in Table ~\ref{tab:5}\\
    HCN & trends as in Table ~\ref{tab:5}\\
    HCO$^+$ & trends as in Table ~\ref{tab:5}\\
    \hline
    \end{tabular}
    \end{center}
\end{table*}

\begin{table*}
    \caption{Temporal trends of molecular fractional abundances, computed for different values of cosmic ray
    ionization rate ($\zeta$). Comparison of hot core chemistry for
    $\zeta$ = 1 and $\zeta$ = 10 while other parameters have the
    standard values (see Table ~\ref{tab:1}). }\label{tab:6}
    \begin{center}
    \begin{tabular}{l l}
    \hline
    Molecule & Response to higher cosmic ray ionization rate\\
    \hline
    CO, H$_2$O & unchanged between $\zeta$ = 1 and $\zeta$ = 10, insensitive tracer of $\zeta$\\
    CS & reduced for $\zeta$ = 10 case by $\approx \times$ 10, good inverse tracer of $\zeta$\\
    OCS & increased for $\zeta$ = 10 by $\approx \times$ 10, linear tracer of $\zeta$\\
    CH$_3$CN & early time ($\approx$ 1 $\times 10^{4}$ yrs) reduced for $\zeta$ = 10,
    but much increased \\
    & at late time ($\approx$ 1 $\times 10^{6}$ yrs), tracer strongly increasing with
    $\zeta$\\
    CH$_3$OH & initially high in both cases, but removed by chemistry \\
    & earlier
    ($\approx$ 1 $\times 10^{5}$ yrs) for $\zeta$ = 10, good inverse tracer of
    $\zeta$\\
    SO & increased significantly for $\zeta$ = 10, good tracer for $\zeta$\\
    SO$_2$ & increased (by $\approx \times$ 10) for $\zeta$ = 10, linear tracer of $\zeta$\\
    H$_2$S & increased significantly for $\zeta$ = 10, good tracer for $\zeta$\\
    H$_2$CS & no significant change, insensitive tracer for $\zeta$\\
    H$_2$CO & reduced for $\zeta$ = 10 after $\approx$ 1 $\times 10^{5}$ yrs, good inverse tracer of $\zeta$\\
    CH$_2$CO & large in both case, unchanged, insensitive tracer for $\zeta$\\
    C$_2$H$_5$OH & undetectable in both cases\\
    HCOOCH$_3$ & barely detectable for $\zeta$ = 1, reduced for $\zeta$ = 10, but inverse tracer of $\zeta$\\
    HNC & increased for $\zeta$ = 10 by $\approx \times$ 10, linear tracer of $\zeta$\\
    HCN & increased significantly for $\zeta$ = 10, good tracer for $\zeta$\\
    HCO$^+$ & not detectable for $\zeta$ = 1, but increased $\approx \times$ 10 for $\zeta$ = 10\\
    \hline
    \end{tabular}
    \end{center}
\end{table*}

\begin{table*}
    \caption{Temporal trends of molecular fractional abundances, computed for different values of initial
    elemental abundance ratios (other parameters have the standard values,
    see Table ~\ref{tab:1}). }\label{tab:7}
    \begin{center}
    \begin{tabular}{l l}
    \hline
    Molecule & Response to changes in initial elemental abundances\\
    \hline
    CO, H$_2$O$^{a}$ & unchanged between models (follows total carbon abundance)\\
    CS & always abundant but removed after $\approx$ 1 $\times
    10^{5}$ yrs from hot cores \\
    & in cases with highest total oxygen abundance to form SO \\
    OCS$^{c}$ & follows total sulfur abundance\\
    CH$_3$CN$^{b}$ & follows total nitrogen abundance, a good discriminant between \\
    & solar and early Universe types \\
    CH$_3$OH & almost unchanged between models, since the abundance \\
    & mainly follows unchanged carbon\\
    SO$^{c}$ & follows total sulfur abundance\\
    SO$_2$$^{c}$ & follows total sulfur abundance\\
    H$_2$S$^{c}$ & not very sensitive to the total sulfur abundance, removed at late
    times ($\approx$ 1 $\times 10^{6}$ yrs) \\
    H$_2$CS$^{c}$ & not very sensitive to the total sulfur abundance, removed at late
    times ($\approx$ 1 $\times 10^{6}$ yrs)  \\
    H$_2$CO & inversely sensitive to total oxygen abundance\\
    CH$_2$CO$^{a}$ & high abundances anti-correlated with total oxygen abundance\\
    C$_2$H$_5$OH & negligible but anti-correlated with total oxygen abundance\\
    HCOOCH$_3$$^{a}$ & low abundances and anti-correlated with total oxygen abundance\\
    HNC$^{b}$ & follows total nitrogen abundance\\
    HCN$^{b}$ & follows total nitrogen abundance\\
    HCO$^+$ & negligible but anti-correlated with total oxygen abundance\\
    \hline
    \end{tabular}
    \end{center}
    $^{a}$ : Hot cores tracers CO, H$_2$O follow the total oxygen
    abundance while CH$_2$CO (and HCOOCH$_3$) are anti-correlated with total oxygen
    abundance; $^{b}$ : HCN, HNC and CH$_3$CN are all potential
    discriminants between solar and early Universe elemental
    abundances through the nitrogen content; $^{c}$ : typical hot
    core molecules as SO, SO$_2$ and OCS are correlated with total
    sulfur abundance while H$_2$S and H$_2$CS are rather insensitive.
\end{table*}

\begin{table*}
    \caption{Detectable tracers of extragalactic hot core chemistry for four examples
    of models likely to be appropriate for representing the four kinds of
    galaxy described in the Sect. ~\ref{sec:gal} and mentioned in the second
    line of this table. The limit of detectability has been taken to be
    (n(X)/n(H)) = 1 $\times 10^{-12}$, as is typical for hot core molecules in the Milky Way.
    Under this limit the species are not detectable (symbol $\times$). Otherwise, they
    are marked with the symbol $\checkmark$.}\label{tab:8}
    \begin{center}
    \begin{tabular}{c c c c c}
    \hline
    & Model 0 & Model 15 & Model 17 & Model 18\\
    \hline
    Type of  & Spiral Normal & Starburst & low-metallicity & high redshift\\
    Galaxy & IC 342, NGC 4736 & M 83 & IC 10 & Cloverleaf QSO, APM 08279\\
    \hline
    CS & $\checkmark$ & $\checkmark$ & $\checkmark$ & $\checkmark$\\
    OCS & $\checkmark$ & $\checkmark$ & $\checkmark$ & $\times$\\
    CH$_3$CN & $\checkmark$ & $\checkmark$ & $\checkmark$ & $\checkmark$\\
    CH$_3$OH & $\checkmark$ & $\checkmark$ & $\checkmark$ &$\times$\\
    SO & $\checkmark$ & $\checkmark$ & $\checkmark$ & $\checkmark$\\
    SO$_2$ & $\checkmark$ & $\checkmark$ & $\checkmark$ & $\times$\\
    H$_2$S & $\checkmark$ & $\checkmark$ & $\checkmark$ & $\times$\\
    H$_2$CS & $\checkmark$ & $\checkmark$ & $\checkmark$ & $\checkmark$\\
    H$_2$CO & $\checkmark$ & $\checkmark$ & $\checkmark$ & $\checkmark$\\
    CH$_2$CO & $\checkmark$ & $\checkmark$ & $\times$ & $\times$\\
    C$_2$H$_5$OH & $\times$ & $\times$ & $\times$ & $\times$\\
    HCOOCH$_3$ & $\checkmark$ & $\checkmark$ & $\times$ & $\times$\\
    HNC & $\checkmark$ & $\checkmark$ & $\checkmark$ & $\checkmark$\\
    HCN & $\checkmark$ & $\checkmark$ & $\checkmark$ &$\checkmark$\\
    HCO$^{+}$ & $\times$ & $\times$ & $\times$ & $\checkmark$\\
    \hline
    \end{tabular}
    \end{center}
\end{table*}

\acknowledgments
\begin{acknowledgements}
    Acknowledgments

Dr. Estelle Bayet acknowledges financial support from a Leverhulme
Trust Research Grant. Dr. Serena Viti acknowledges individual
financial support from a PPARC Advanced Fellowship. We thank the
referee for a careful reading and detailed suggestions that helped to
improve the original version of the paper.
\end{acknowledgements}

\bibliographystyle{aa}

\bibliography{references}

\end{document}